\documentclass[aps,prl,twocolumn,showpacs,superscriptaddress]{revtex4}
\usepackage{graphicx}
\usepackage{dcolumn}
\usepackage{bm}

\begin{document}

\title{Correlations of mutual positions of charge density waves nodes in side-by-side placed InAs wires measured with scanning gate microscopy.}

\author{A.A.~Zhukov}
\affiliation{Institute of Solid State Physics, Russian Academy of
Science, Chernogolovka, 142432 Russia}

\author {Ch.~Volk}
\author {A.~Winden}
\author {H.~Hardtdegen}
\affiliation{Peter Gr\"unberg Institut (PGI-9), Forschungszentrum
J\"ulich, 52425 J\"ulich, Germany} \affiliation{ JARA-Fundamentals of
Future Information Technology, Forschungszentrum J\"ulich, 52425
J\"ulich, Germany}
\author {Th.~Sch\"apers}
\affiliation{Peter Gr\"unberg Institut (PGI-9), Forschungszentrum
J\"ulich, 52425 J\"ulich, Germany} \affiliation{ JARA-Fundamentals of
Future Information Technology, Forschungszentrum J\"ulich, 52425
J\"ulich, Germany} \affiliation{II. Physikalisches Institut, RWTH
Aachen University, 52056 Aachen, Germany}
\date{\today}

\begin{abstract}
We investigate the correlations of mutual positions of charge density waves nodes in side-by-side placed InAs nanowires in presence of a conductive atomic force microscope tip served as a mobile gate at helium temperatures. Scanning gate microscopy scans demonstrate mutual correlation of positions of charge density waves nodes of two wires. A general mutual shift of the nodes positions and ``crystal lattice mismatch'' defect were observed. These observations demonstrate the crucial role of Coulomb interaction in formation of charge density waves in InAs nanowires.
\end{abstract}
\pacs{73.23.Hk, 73.40.Gk, 73.63.Nm}

\maketitle

Essential efforts were made towards investigations of electronic transport in InAs \cite{Hansen2005, Ford2009, Scheffler2009, Dhara2009, Hernandez2010, Bloemers2011, Wirths2011, Liang2012}, InN \cite{Richter2008}, core/shell GaAs/InAs \cite{Gul2014} and Si/Ge \cite{Hao2010} nanowires made by bottom-up approach partially because of the rich physics demonstrated by such structures and because of the possible applications of them in electronics and quantum computing. The most experiments done on wires with normal-metallic contacts are dedicated to study magneto-transport properties of field-effect transistors made of nanowires using n-type doped Si substrate as a back gate \cite{Hansen2005, Ford2009, Scheffler2009, Dhara2009, Hernandez2010, Bloemers2011, Wirths2011, Richter2008, Gul2014, Hao2010} or PEO/LiClO solid electrolyte  as a surrounded gate \cite{Liang2012} to alter the number of electrons or holes in the system. These types of measurements reveal the response of the nanowire as the whole masking the local peculiarities of the systems under investigation.

The so called scanning gate microscopy (SGM) technique used the conductive atomic-force microscope tip as a mobile gate became the standard one and helps to reveal the local electronic transport properties of quantum dots \cite{Pioda2004, Gildemeister2007}, quantum point contacts \cite{TopinkaS2000, TopinkaN2001, SchnezQPC2011} or quantum rings \cite{Hackens2006} based on  two-dimensional electron gas or quantum dots based on one-dimensional objects such as nanowires \cite{Bleszynski2005, Zhukov2012, Zhukov2012SW, Zhukov2014SWComp, Zhukov2014} and nanotubes \cite{Bockrat2001, Woodside2002, Zhukov2009}. The last experiments done on high quality InAs nanowires show a presence of charge density wave (CDW) with wavelength ranges from two-hundred nanometers to one micron formed by electrons of top subband of InAs nanowire  \cite{Zhukov2012SW}. Detailed SGM experiments help to define three groups of electrons in InAs nanowire \cite{Zhukov2014SWComp}: the electrons of top subband, the electrons belonging to ``disordered sea'' and electrons  localized on surface defect centers. These experiments revealed robustness of such charge density wave to application of high source-to-drain voltage, possible coexistence of charge density waves with 2 and 3 nodes and, finally, the splitting of the maxima of charge density wave \cite{Zhukov2014SWComp}. 

Three possible mechanisms of charge density wave formation are considered. The first one is a simple standing waves model \cite{Zhukov2012SW} excluding formally the influence of charge of the electrons itself on position and wavelength of charge density wave. More cumbersome the second and the third ones are Friedel oscillations and Wigner crystallization formed by Luttinger liquid of electrons in InAs nanowire \cite{Meyer2006, Gindikin2007, Traverso2012, Traverso2013}. None of them appear to be suitable to explain charge density wave with micron wavelength \cite{Zhukov2014SWComp}, but, nevertheless, all three look quite applicable for explanation of charge density wave with wavelengths of hundreds microns. The last experiment \cite{Zhukov2014} demonstrates the presence of slalom-like stable patterns in SGM images interpreted as Wigner crystal instability \cite{Meyer2006}. This experiment pointed on the principal importance of charge of electrons of top subband on wavelength and charge density wave positions, but more solid evidences look to be necessary to exclude simple standing waves model from consideration.

In the current paper we present a set of experiments made on side-by-side placed InAs nanowires using SGM technique. Observable correlations of mutual positions of charge density wave nodes in neighbor wires allow us to reveal the crucial role of charge of electrons in formation of CDW. No essential correlation is found for electrons in ``disordered sea'' at high electron density (high back gate voltages $V_{BG}=11.0-12.0$~V).  

In our experiment we study a nominally undoped InAs nanowires grown by
selective-area metal-organic vapor-phase epitaxy \cite{Akabori2009}. The
diameter of the wires is 100\,nm. The wires were placed on an $n$-type
doped Si (100) substrate covered by a 100~nm thick SiO$_2$ insulating
layer. The n-type doped Si(100) substrate served as the back-gate electrode. The evaporated Ti/Au contacts to the wire as well as the markers of the
search pattern were defined by electron-beam lithography. The distance
between the contacts is $L_{wire}=3.0$~$\mu$m. A scanning electron
micrograph of the sample is shown in Fig.~1(a). The
source and drain metallic electrodes connected to the wires are marked
by 's' and 'd'. We define the ``upper'' wire as the one which is shorter on drain side, the other one is defined as the ``lower'' wire.

\begin{figure*}
\includegraphics[width=1.8\columnwidth]{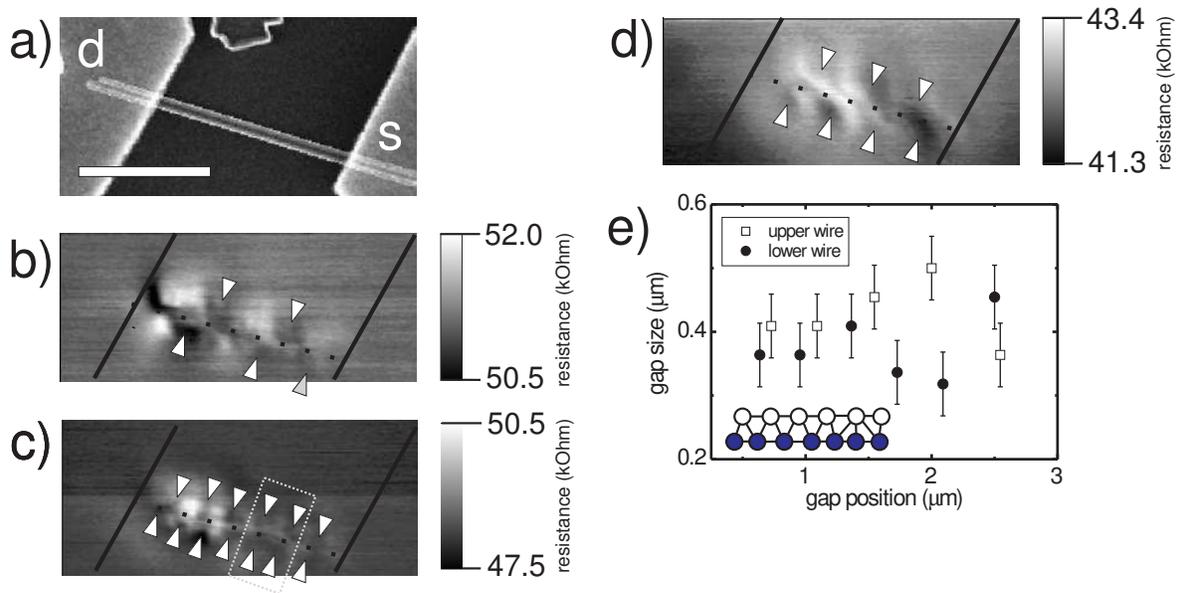}
\caption{a) Scanning electron micrograph of the InAs side-by-side placed wires with metallic contacts. The source and drain contact pads are marked by 's' and 'd'. We define the ``upper'' wire as the one which is shorter on drain side, the other one we define as the ``lower''. The horizontal scale bar corresponds to 2~$\mu$m. All scans a) to d) have the same scale. b) - d) are scanning gate microscopy scans made with a tip voltage $V_t=0$~V and a back gate voltage of $V_{BG}=-0.2$~V, $+0.8$~V and $+3.5$~V, correspondently. The tip-to-surface distance is maintained at 300~nm. White triangles mark the nodes of charge density waves. The gray triangle in Fig.~b) notes the beginning of formation of the third node in lower wire. Solid lines and dotted line mark the edges of the metallic contacts and the position of the nanowires. e) shows the dependence of the gaps sizes vs. gaps positions for upper (open squares) and lower (solid circles) wires. The bottom inset is a general scheme of formation of ``lattice mismatch '' defect in accordance with measured gaps (distances between nodes) sizes. The drain contact position is denoted as ``0''~$\mu$m on abscissa in Fig.~e).}
\label{Fig1}
\end{figure*}

All measurements were performed at $T=4.2$~K. The conductive tip of a
home-built scanning probe microscope \cite{AFM} is used as a mobile
gate during scanning gate imaging measurements. We keep the tip 300~nm
above the SiO$_2$ surface for ``far-placed-tip'' measurements and 200~nm for ``close-placed-tip'' ones similar to \cite{Zhukov2014SWComp}. All scanning gate measurements are performed keeping the potential of the scanning probe microscope tip ($V_t$) as well as the back-gate voltage ($V_{BG}$) constant. The electrical circuit of the scanning gate imaging measurements can be found elsewhere \cite{Zhukov2009}. The conductance of the wire during the scan is measured in a two-terminal circuit by using a standard lock-in technique. A driving AC current with an amplitude of $I_{AC}=1$~nA at a frequency of 231~Hz is applied while the voltage is measured by a differential voltage amplifier.

Figures~1b)~-~1d) show the scanning gate microscopy scans obtained in ``far-placed-tip'' regime. In this regime the top subband electrons essentially alter the resistance of wires \cite{Zhukov2014SWComp}. Tip 
voltage is kept at $V_t=0$~V, while the back gate voltage is set to 
$V_{BG}=-0.2$~V, $+0.8$~V and $+3.5$~V for scans b) to d), correspondently.
Arrows mark positions of the nodes of charge density waves. 

Fig.~1b) demonstrates two-nodes charge density wave as for the upper so for the lower wires. The wavelength or node-to-node distance is the same for both wires and equal to 1.1~$\mu$m. Mutual shift of nodes position for upper and lower wires of $0.25-0.30$~$\mu$m is observed as well. The beginning of formation of the third node for lower wire is fairy visible and its position is marked by a gray triangle.

Fig.~1c) shows well developed charge density waves with six nodes for the upper wire and seven nodes for the lower one. The general mutual shift of the charge density wave nodes positions for each wire is presented as well as the clearly visible deformation (non-equidistant positions of nodes) of both charge density waves. The region of the this deformation is marked with gray rectangle.

Fig.~1d) demonstrates three-nodes charge density wave for the upper wire with  wavelength of $~0.77$~$\mu$m and four-nodes charge density wave for the lower wire with wavelength of $~0.67$~$\mu$m. It is possible to note that the second node of charge density wave of the upper wire is placed exactly in between the second and the third nodes of the lower wire. No deformation of the charge density waves is observed.

For better visualization of defect shown in Fig.~1c) the sizes of the gaps (distances between charge density wave nodes) and their positions are presented in Fig.~1e). Error bars of 50~nm mark the precision of nodes allocation. A well defined ``defect'' is located around position of 2~$\mu$m. Here ``0''~$\mu$m denotes the drain contact position. 

Figs.~2a) to f) show scanning gate microscopy measurements obtained in ``close-placed-tip'' regime. For this kind of measurements tip is able to influence locally on electrons of ``disordered sea'' \cite{Zhukov2014SWComp}. Measurements are done keeping tip voltage at $V_t=0$~V and back gate voltage swept from $V_{BG}=11.0$~V in Fig.~2a) to $V_{BG}=12.0$~V in Fig.~2f) in steps of 0.2~V. No clear correlations or pronounced short scale ($<150$~nm) defects or weak points in wires could be detected on these scans.

\begin{figure*}
\includegraphics [width=1.5\columnwidth]{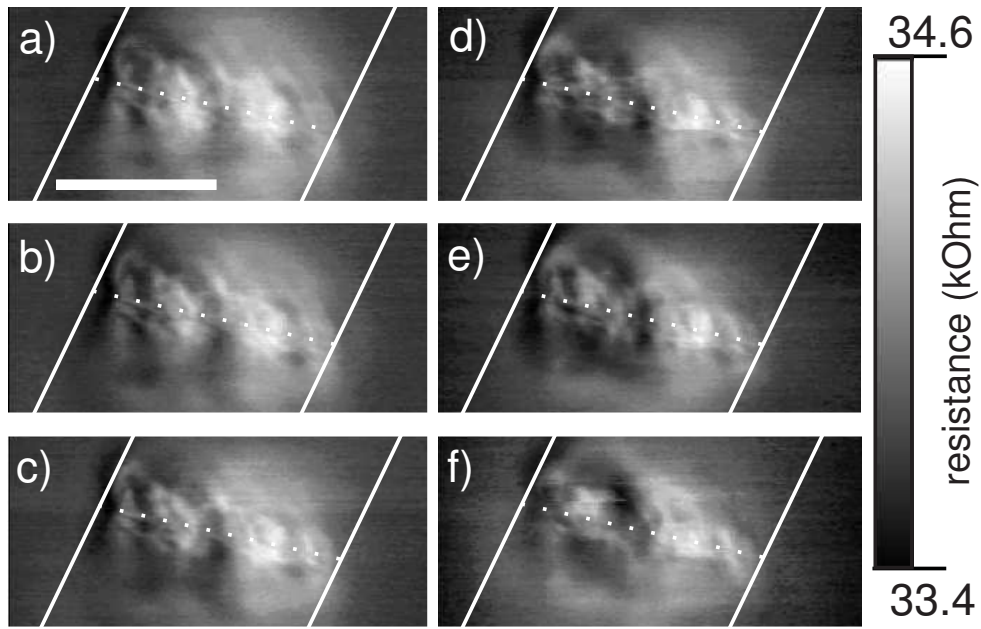}
\caption{Figs.~a) to f) show scanning gate microscopy scans of wires at high electron density. The tip-to-surface distance is 200~nm. No well defined correlations for electrons of the ``disordered sea'' can be observed.
Solid lines and dotted line mark the edges of the metallic contacts and the position of the nanowires, correspondently. The horizontal scale bar in Fig. a) corresponds to 2~$\mu$m, this scale is the same for all figures. The tip voltage $V_t=0$~V is kept the same for all scans. The back gate voltage is increased from 11.0~V for Fig.~a)  to 12.0~V for Fig.~f) in steps of 0.2~V.}
\label{Fig2}
\end{figure*}

Lets discuss the experimental results in more details. The high quality of the nanowires allows us to perform measurements not far from the pinch-off back gate voltage $V_{BG pinch-off} \sim -1.5$~V. So, even at $V_{BG} \sim 0$~V no sign of the Coulomb blockade circles \cite{Bleszynski2005, Zhukov2012} because of the internal weak links or because of low quality metallic contacts is observed.

The three scans presented in Figs.~1b) - d) show that the upper wire tends to contain one or two electrons less in the top subband than the lower one. It means the presence of constant misalignment of the bottoms of the conductive bands for two wires although the length of them defined by metallic contact pads are actually the same.

For Fig.~1b) the mutual shift of the charge density waves nodes positions for $250 - 300$~nm has a quite simple interpretation. This might happened because the Coulomb repulsion of CDW nodes, the Coulomb energy for the electrons placed 250~nm apart is of $V_C \approx 0.16$~eV and becomes comparable to the bath temperature of $T=4.2$~K \cite{Zhukov2014SWComp}. 



It is worth noting almost the perfect self-alignment of the three-nodes charge density wave and four-nodes one in Fig.~1d). The small number of nodes allows to settle both charge density waves symmetrically without essential deformations despite the difference of the wavelengths of 0.77~$\mu$m for the upper wire and 0.67~$\mu$m for the lower CDW. Symmetry of nodes positioning means the similar sizes of close to metallic contacts depletion regions for both wires.   

The most striking plot is shown in Fig.~1c). This figure is the main result of the current paper. As in the upper wire so in the lower wire well defined charge density waves are formed. The lengths of them are the same. They are defined by the positions of the contact pads as well as by the near by placed depletion regions of 200-300~nm \cite{Bleszynski2005, Zhukov2012}. So, there are two ``crystals'' of the same length but with six ``atoms'' in the upper wire and seven ones in the lower. Mutual deformation because of their ``lattice mismatch'' is observed around the position at 2~$\mu$m (see Fig.~1e)). Its position is marked on the scan in Fig.~1c) by a gray rectangle. The observable mutual deformation of two ``crystal lattices'' might not cause any surprise because the size of energy of Coulomb interactions of electrons of these ``crystals'' is of the same order of magnitude as the bath temperature of experiment, as mentioned previously.

Thus, the experimental data presented in Fig.~1c) demonstrates clearly the crucial role of the Coulomb interaction for the definition as the wavelengths as well as the node positions of the charge density waves formed by electrons of the top subband in InAs wires. Hence, the simple model of ``standing waves'' \cite{Zhukov2012SW} should be excluded from consideration, and the remaining two, namely, the Friedel oscillations and Wigner crystal formation models might be considered in further investigations for small wavelengths ($250-300$~nm) CDW.

Additionally, the presented experimental data is an additional support of interpretation of the previous experimental results (see Fig.~4 in \cite{Zhukov2014} where slalom-like patterns were explained as possible formation of zig-zag Wigner crystal \cite{Meyer2006}.

We should point on fact that besides essential progress have been made in measurements using lithography defined local gates \cite{Fasth} the results presented in current paper could be obtained with SGM technique only.

As it was noted above, no clear mutual correlations for electrons of ``disordered sea'' or pronounced short scale ($<150$~nm) defects in wires could be detected on scans presented in Fig.~2. These scans also confirm that generally a smooth potential background is observed in our system not only at low back gate voltages but at high back gate voltages as well.

In conclusion, we performed for the first time scanning gate microscopy measurements on two InAs quantum wires without internal tunneling barriers placed side-by-side. Measurements are done in two regimes: the tip is 300~nm above the surface in ``far-placed-tip'' regime and the tip is placed at the height of 200~nm for ''close-placed-tip'' one. In ''far-placed-tip'' regime the  correlations of positions of charge density waves nodes for two wires, namely, the general mutual shift and ``crystal lattice mismatch'' defect are observed for the first time. No pronounced correlations are found for electrons of the ``disordered sea'' in the ``close-placed-tip'' regime for high back gate voltages. 

This work is supported by the Russian Foundation for Basic Research (Grants RFBR 14-02-00192a and RFBR 13-02-12127ofi-m),
programs of the Russian Academy of Science, the Program for Support of
Leading scientific Schools, and by the International Bureau of the
German Federal Ministry of Education and Research within the project
RUS 09/052.

{}

\end{document}